# Damage detection of high-speed railway box girder using train-induced dynamic responses

Xin Wang[a], Yi Zhuo[b] and Shunlong Li[*a]

[a] *School of Transportation Science and Engineering, Harbin Institute of Technology, Harbin 150090, China*

[b] *China Railway Design Corporation, 300142 Tianjin, China*

**Abstract.** This paper proposes a damage detection method based on the train-induced responses of high-speed railway box girder. Under the coupling effects of bending and torsion, the traditional damage detection method based on the Euler beam theory cannot be applied. In this research, the box girder section is divided into different components based on the plate element analysis method. The strain responses were preprocessed based on Principal Component Analysis (PCA) method to remove the influence of train operation variation. The residual error of autoregressive (AR) model was used as a potential index of damage features. The optimal order of the model was determined based on Bayesian information criterion (BIC) criterion. Finally, the confidence boundary (CB) of damage features (DF) constituting outliers can be estimated by Gaussian inverse cumulative distribution function (ICDF). The numerical simulation results show that the proposed method in this paper can effectively identify, locate and quantify the damage, which verifies the accuracy of the proposed method. The proposed method effectively identifies the early damage of all components on the key section by using four strain sensors, and it is helpful for developing effective maintenance strategies for high-speed railway box girder.

**Keywords:** damage detection; plate element analysis method; confidence boundary; high-speed railway box girder; maintenance strategies.

## 1. Introduction

The total operating mileage of high-speed railways has exceeded 35,000 km in China, including high-speed railway bridges of 16,000 km. More than 85% of the high-speed railway bridges are constructed with prestressed concrete simply supported box girders [1]. High-speed railway bridges face various potential risks during service, such as natural disasters, fatigue, corrosion, etc [2]. These potential problems will lead to different degrees of damage, which may affect the operational performance and safety of these bridges. Therefore, it is essential to maintain these railway bridges regularly through effective monitoring strategies. Thus, early damage detection of bridges has become an important and indispensable part of structural health monitoring (SHM) systems for high-speed railways, the application of new methods or new materials in the field of SHM has been widely studied [3-8]. Extensive research efforts have been devoted to damage detection and many effective methods have been proposed [9-14].

These methods can be divided into model updating or data-driven methods [15]. Model updating modifies FEM through experimental data, which is not suitable for real-time SHM for large bridges due to complex calculation. The data-driven methods extract valuable information from time series data acquired in the field, which is no need for any structural analysis modeling or updating of the FEM, and online damage detection can be realized through data mining technology [16]. Since the modal characteristics are directly related to the structure stiffness, the structure stiffness is expected to change in the presence of damage. Therefore, modal-based damage identification method is the most common [17-19]. As the method based on Operational Modal Analysis (OMA) requires recognition of high order mode shapes, it is considered insensitive to early damage [20].

---

*Corresponding author, Professor, E-mail: lishunlong@hit.edu.cn

Many techniques have been successfully applied to the extraction of structural damage-sensitive features, such as symbolic data [21], wavelet components [22] and basic signal statistics. However, since the parameters of the autoregressive (AR) model reflect the inherent characteristics of the structures, model coefficients or residual errors can be extracted through time series analysis as damage sensitive features [23]. In addition, AR model only depends on the response of structure, so it is widely used in the field of damage identification. Autoregressive with exogenous input (ARX) [24], autoregression and autoregression of exogenous input (ARARX) [25], autoregressive moving average (ARMA) [26], autoregressive moving average and exogenous input (ARMAX) [27] are also used for feature extraction as time series models.

The damage detection techniques generally distinguish structural responses under environmental and operational variations (EOV) from those induced by damage through feature modeling [28]. Regression based methods (such as multiple linear regression [29, 30]) or only based on latent variables (such as principal component analysis [18, 20]) are widely used in feature modeling. For the data-driven methods, feature discrimination aims at classifying the features into healthy or damaged by supervised or unsupervised learning algorithms, where statistical process control [22] or MLP neural network [31] belong to supervised learning algorithms. Since there is little or no data obtained from damaged structures, unsupervised learning algorithms such as novelty detection methods have been widely studied and applied [32]. As a novelty detection technology, outlier analysis can identify whether the structure is damaged by fitting the data under baseline conditions with probability distribution and then testing whether the new data conforms to the same distribution [33].

Although most SHM methods use environmental excitation or free vibration signals for analysis, some studies use vehicle-induced response for bridge damage identification. Under unknown moving load, Cavadas et al. [29] collected displacement and rotation data of beam frame. By defining baseline data, outlier analysis is used to identify data beyond the baseline range as damaged data. The method successfully detected a 20% stiffness reduction of a beam element with a length of 30cm, but its limitation is that only a single load is considered. Gonzalez and Karoumi [33] used the machine learning model based on ANN and Gaussian processes to train the deck accelerations and bridge weigh-in-motion data, so as to classify the data as healthy or damaged. Nie et al. [34] proposed a damage detection method using two sensors. The local damage index is defined using the cross-correlation coefficient between the measured responses, and the effectiveness of the proposed method is verified by the simply supported beam experiment in the laboratory. Although this field has been extensively studied, there are still few cases of damage detection based on train-induced dynamic response. The environmental and operational variations in structural response are often ignored in most damage detection methods, the damage types of structure are limited, the load excitation is very specific, and the vehicle-bridge coupling effect is rarely considered. The bridges are simply modeled as Euler-Bernoulli beams, which limits their usability in real and complex bridges. In this study, when a train passes through the 32m simply supported box girder with two lanes, the train load is distributed eccentrically. Under the action of bending-torsion coupling effect, the deformation of box girder does not conform to the assumption of plane section, and it cannot be simplified as an Euler-Bernoulli beam model [1]. Therefore, the FEM is modeled using solid element, a multi-body dynamics model of train-track-bridge system is established, the strain response of different components of each key section is extracted by simulating the operation of the high-speed train.

This paper proposes a damage detection method based on the train-induced responses of high-speed railway box girder. Under the coupling effects of bending and torsion, the traditional damage detection method based on the Euler beam theory cannot be applied. In this research, the box girder section is divided into different components based on the plate element analysis method. The FEM verified by experiments was used to simulate the baseline and damage conditions. In addition, the effects of noise, different train speeds, different train weights and vertical track instability are considered under all conditions. The numerical simulation results show that the proposed method in this paper can effectively identify, locate and quantify the damage, which verifies the accuracy of the proposed method.

The rest of paper is organized as follows. Section 2 firstly introduces the method of removing the principal component of train-induced responses, then describes the residual error of AR model is used as damage feature in feature extraction, and finally introduces the outlier analysis based on a newly proposed damage index. Section 3 presents the numerical simulation of a typical 32 m simply supported box girder of high-speed railway, which verifies the accuracy and robustness of the proposed method. Some conclusions are finally drawn in Section 4.

## 2. Damage detection methodology

### 2.1 Analysis on the removal of principal components

In order to separate EOVs from dynamic responses containing train-induced signals and obtain the main damage-sensitive features, principal component analysis [18] was used in this paper for feature modeling, and some related studies have shown that PCA can effectively remove the effects of EOVs [35,36]

The PCA model can be written as:

$$Y_{k \times n} = X_{k \times n} \cdot T_{n \times n} \quad (1)$$

where $X$ is the response time-series data, $k$ is the number of passing trains, and $n$ is the length of strain signal, $T$ is an $n$-by-$n$ orthonormal linear transformation matrix.

The covariance matrix of the $X$ in the baseline condition, **C**, which can be expressed as:

$$C = T \cdot \Lambda \cdot T \quad (2)$$

where $T$ and $\Lambda$ are matrixes obtained by the singular value decomposition of the covariance matrix of $X$.

Since the purpose of this research is to detect damage with local characters, the feature modelling procedure involves removing the most significant principal components from the features and retaining the remaining principal components for subsequent statistical analysis. The matrix $\Lambda$ can be divided into a matrix with the first $p$ eigenvalues and a matrix with the remaining $m$-$p$ eigenvalues. In this study, the value of $p$ is determined based on the rule of thumb in which the cumulative percentage of the variance reaches 80% **[9]**.

After the value of $p$ is determined, the first $p$ component of matrix $Y$ can be remapped to the original space using the following formula:

$$\begin{aligned} X_p &= X \cdot \hat{T} \cdot \hat{T}^T \\ \tilde{X} &= X - X_p \end{aligned} \quad (3)$$

where $\hat{T}$ is the first $p$ columns of $T$, and $\tilde{X}$ is the strain time domain signals after removing the principal component $X_p$. After preprocessing the signals of all conditions, **Eq. (4)** can be obtained:

$$\begin{aligned} \tilde{X}_{baseline} &= X_{baseline} - X_{p\_baseline} \\ \tilde{X}_{damaged} &= X_{damaged} - X_{p\_baseline} \end{aligned} \quad (4)$$

## 2.2 Feature extraction

In the field of Structural Health Monitoring (SHM), time series analysis attempts to fit mathematical models with time series data, and the AR models are widely used to extract damage sensitive features of structures **[37]**.

The AR($m$) model can be written as:

$$x_j = \sum_{i=1}^{m} a_i x_{j-i} + \varepsilon_j \quad (5)$$

where $x_j, j = m+1, m+2, m+3, \ldots, n$ is the response time-series data from $\tilde{X}$ in **Eq. (3)**, $m$ is the number of AR parameters in **Eq. (5)**, $\varepsilon_j$ is the residual error of the signal $x_j$.

According to **Eq. (5)**, the matrix form of $n$-$m$ algebraic equations can be expressed as:

$$S = H \cdot w + \varepsilon$$
$$\Downarrow$$
$$\begin{bmatrix} x_{m+1} \\ x_{m+2} \\ \vdots \\ x_n \end{bmatrix} = \begin{bmatrix} x_1 & x_2 & \cdots & x_m \\ x_2 & x_3 & \cdots & x_{m+1} \\ \vdots & \vdots & \ddots & \vdots \\ x_{n-m} & x_{n-m+1} & \cdots & x_{n-1} \end{bmatrix} \begin{bmatrix} a_m \\ a_{m-1} \\ \vdots \\ a_1 \end{bmatrix} + \begin{bmatrix} \varepsilon_{m+1} \\ \varepsilon_{m+2} \\ \vdots \\ \varepsilon_n \end{bmatrix} \quad (6)$$

In undamaged and damaged conditions, AR parameters were extracted from the time domain signals after removing the principal component for each sensor respectively, and then the vector of model residuals under the condition of baseline and damage state is used as damage characteristics.

$$\hat{\varepsilon} = S - \hat{S} = S - H \cdot \hat{w} \tag{7}$$

## 2.3 Outlier analysis based on a newly proposed damage index

In this study, a new index has been proposed to indicate the damage characteristics of box girder structures. Damage Features (DF) is defined as the difference between fit ratios (FR), as shown in Eq (8) and (9):

$$FR_1 = \frac{\|s_{measured\_baseline} - \hat{s}_{reconstructed\_baseline}\|}{\|s_{measured\_baseline}\|} \quad FR_2 = \frac{\|s_{measured\_damaged} - \hat{s}_{reconstructed\_damaged}\|}{\|s_{measured\_damaged}\|} \tag{8}$$

$$DF = \frac{|FR_2 - FR_1|}{FR_2} \times 100 \tag{9}$$

where $s_{measured\_baseline}$, $\hat{s}_{reconstructed\_baseline}$ are the measured and reconstructed output from the data set for baseline conditions, respectively. In addition, $s_{measured\_damaged}$, $\hat{s}_{reconstructed\_damaged}$ are the measured and reconstructed output from the data set for damaged conditions, respectively.

The DF values are linear transformations of the residuals, which is approximately subject to normal distribution, and outlier analysis can be performed based on statistical threshold [38, 39]. Under this assumption, the confidence boundary (CB) of DF constituting outliers can be estimated by Gaussian inverse cumulative distribution function (ICDF). The average value of baseline feature vector is $\mu$, the standard deviation is $\sigma$ and the significance level is $\alpha$. The inverse function can be defined according to Gaussian cumulative distribution function (CDF) as follows:

$$CB = invF(1-\alpha) \tag{10}$$

where

$$F(x/\mu,\sigma) = \frac{1}{\sigma\sqrt{2\pi}} \int_{-\alpha}^{x} e^{-\frac{1}{2}\left(\frac{x-\mu}{\sigma}\right)^2} dy, \quad for \quad x \in \mathbb{R} \tag{11}$$

Therefore, when DF of a feature is equal to or greater than CB, the feature can be considered an outlier.

## 3. Numerical simulation and validation of the proposed algorithm

### 3.1 FEM of the widely used simply supported box girder

For the 32m simply supported box girder widely used in high-speed railway, Wang et al. [1] established its three-dimensional (3D) FEM with ANSYS software, and described the modeling process of each component in detail, including the main girder, prestressed steel, track plate, etc. In their study, the FEM is updated according to the measured data of the dynamic load test. The 3D FEM was established as shown in **Fig. 2**. according to the design drawings of the box girder shown in **Fig. 1**.

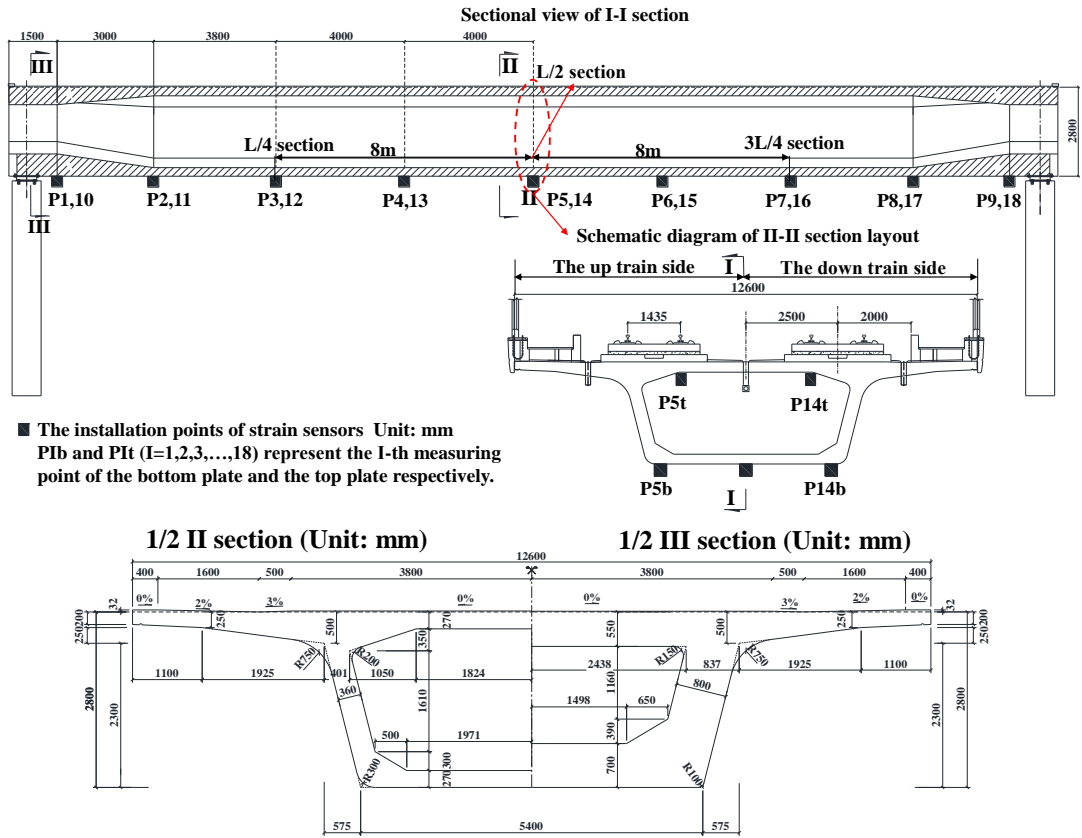

**Fig. 1.** Design drawings and sensor layout of the investigated 32 m box girder.

Universal Mechanism (UM) is a large general-purpose multi-body system dynamics simulation software, which has been widely used in railway transportation, highway transportation, aerospace and other fields. Wang et al. **[1]** imported the ANSYS FEM into UM software and established a multi-body dynamics model of train-track-bridge system of high-speed railway. They imposed stiffness constraints on the bridge in UM to simulate the bearings. The track was modeled as continuous elastic foundation beam, and the foundation below the rails was regarded as a connection of parallel and series linear spring damping system in the vertical and transverse directions in UM. As shown in **Fig. 2**, the track model is established in ANSYS according to the design parameters of high-speed railway track

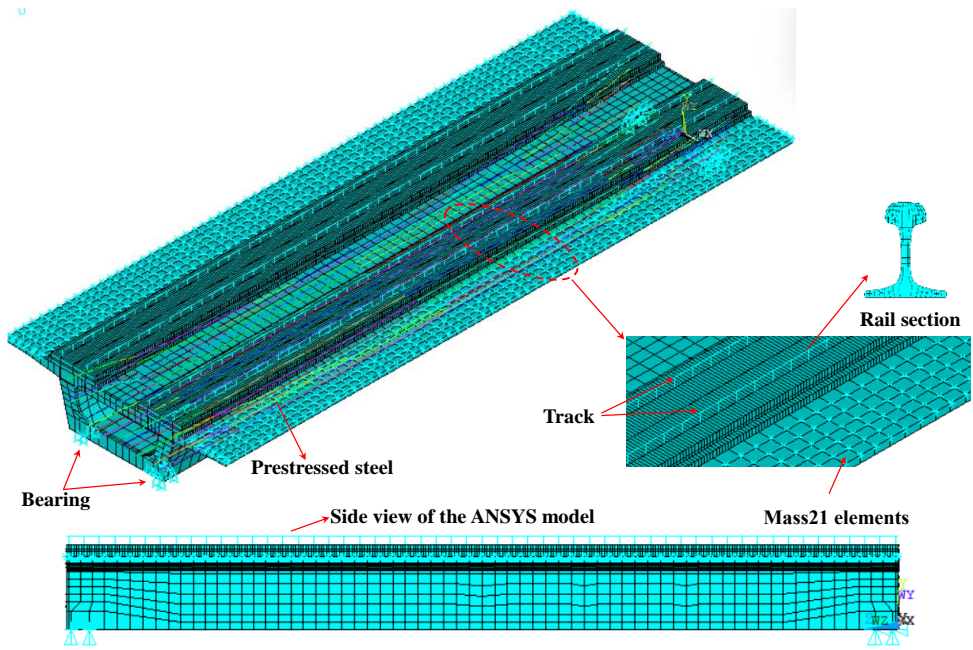

**Fig. 2.** Various views of three-dimensional FEM in ANSYS.

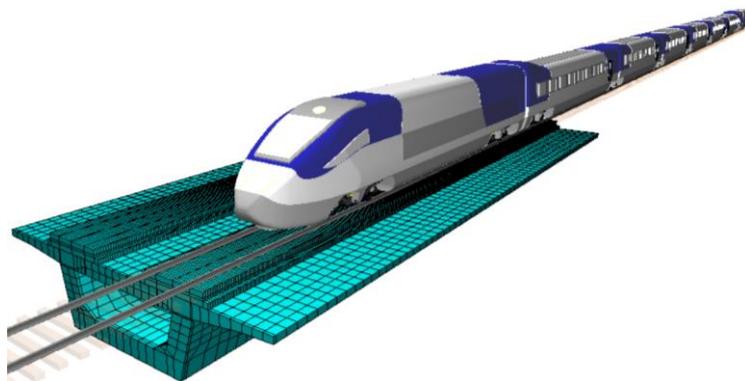

**Fig. 3.** Schematic diagram of multi-body dynamics model in UM.

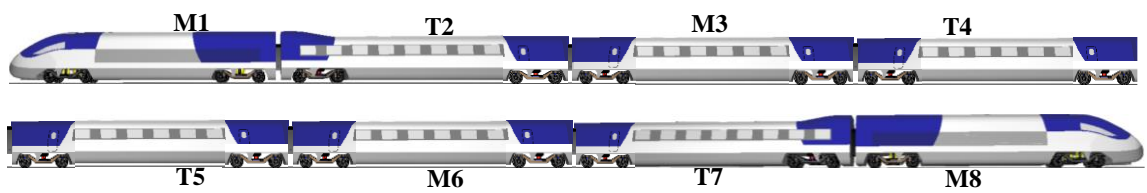

**Fig. 4.** Schematic diagram of the rigid body model of the CRH380 high-speed train.

Table 1. Detailed parameters of high-speed trains

| Train number | Train quality (kN) | Number of passengers | Total weight (kN) | Average axle weight (kN) |
|---|---|---|---|---|
| M1 | 548.8 | 33 | 574.3 | 143.6 |
| T2 | 599.8 | 85 | 666.4 | 166.6 |
| M3 | 585.1 | 85 | 651.7 | 162.9 |
| T4 | 541.0 | 75 | 599.8 | 149.9 |
| T5 | 563.5 | 63 | 612.5 | 153.1 |
| M6 | 599.8 | 85 | 666.4 | 166.6 |
| T7 | 588.0 | 85 | 654.6 | 163.7 |
| M8 | 536.1 | 45 | 571.3 | 142.8 |

The train body, bogie and the wheelsets of the CRH380 were modelled as rigid bodies in accordance with China's High-Speed Railway Code in UM, which were connected with each other through the primary and secondary suspension system as described by Wang et al. [1]. The high-speed train consists 4 motor carriages and 4 trailers, which are simplified as M and T, respectively, in **Fig. 4**. The detailed parameters of high-speed trains are shown in **Table 1**. Under full load condition of the high-speed train, it can carry 556 passengers with a standard weight of 80kg/person. **Fig. 3** shows a schematic diagram of the multi-body dynamics model in UM, when the CRH380 high-speed train passes through the investigated 32 m box girder.

### 3.2 Realistic simulation of damage conditions

Since it is not possible to simulate damage conditions through field experiments, numerical simulations of healthy (baseline) and damage conditions were carried out to verify the method proposed in this paper. After verifying the method, it can be directly applied to the field monitoring data of high-speed railway box girder. **Fig. 5** shows the various combinations of conditions at baseline and damage state.

As shown in **Fig. 5**, supposing that CRH380 high-speed train passes through the investigated box girder at different speeds in UM software for numerical simulation. For the baseline state, 4 different train weights conditions are set in UM, 8 different vertical irregularity spectrums of the track are considered. The UIC good and UIC bad in **Fig. 5** represent the low-interference spectrum and high-interference spectrum respectively, which are widely used in high-speed railways and ordinary railways in Europe [40]. For the baseline state, a total of 96 different conditions were simulated. Three sections, L/4, L/2 and 3L/4, were used to simulate the damaged conditions. Each section was divided into six parts according to the criteria of plate element analysis [1] as shown in **Fig. 6**. The damage of each part was simulated as stiffness reduction. A total of 216 damage conditions were simulated. At the same time, the strain responses of points 1-18 of the box girder in **Fig. 1** are extracted as the data set for the subsequent damage identification, and the sampling frequency is set to 1000 Hz in the numerical simulation.

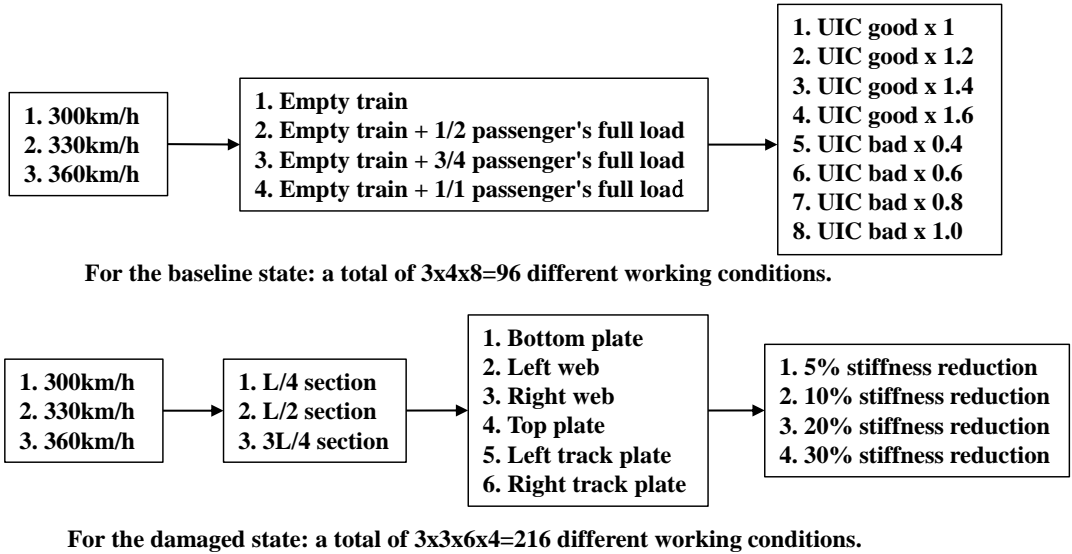

**For the baseline state: a total of 3x4x8=96 different working conditions.**

**For the damaged state: a total of 3x3x6x4=216 different working conditions.**

**Fig. 5.** Various combinations of conditions at baseline and damage state

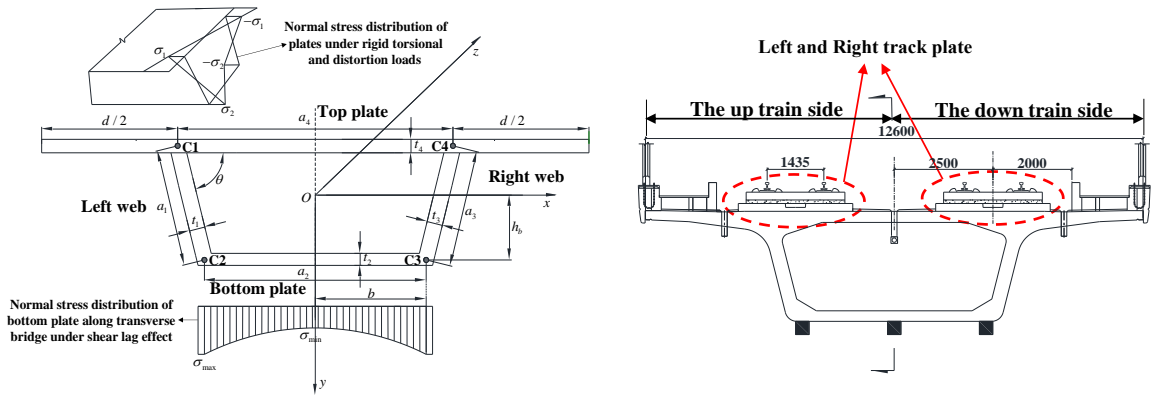

**Fig. 6.** The section is divided into different components based on the plate element analysis method

## 3.3 Verification of proposed method

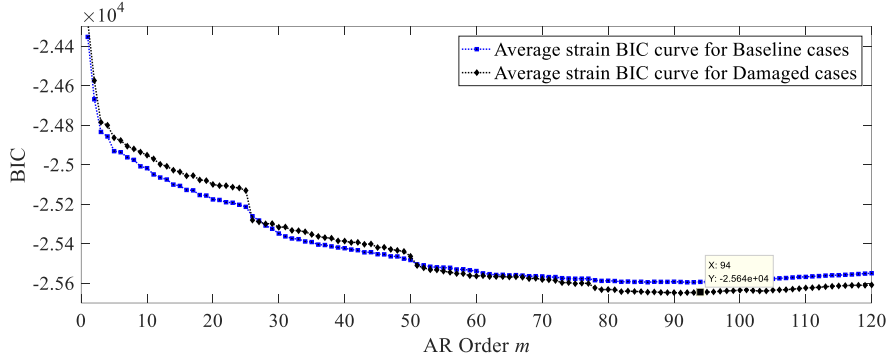

**Fig. 7.** Average strain BIC curve for all cases.

In this study, since the cumulative variance percentage of the first two principal components under different baseline data sets is greater than 80%, the first two principal components are removed in the modeling process of **Section 2.1** to preprocess the data for subsequent analysis of damage identification.

In time series analysis and feature extraction, it is very important to determine the order of AR model. Unreasonable order will lead to insufficient or overfitting of the model. Bayesian information criteria (BIC) **[41, 42]** is a well-known technique for determining the order of AR model. BIC avoids overfitting the model by introducing penalty term. In this paper, BIC is used to obtain the order of AR model. The BIC equation can be expressed as**:**

$$BIC = n\ln(\hat{\boldsymbol{\varepsilon}}^2) + m\ln(n) \tag{12}$$

where $n$ is the number of elements in $\hat{\boldsymbol{\varepsilon}}$ of **Eq. (7)**. The $m_{opt}$ corresponding to the minimum value of BIC is used as the optimal AR order.

For the baseline and damage data sets, each sensor would get a BIC curve for each train passing condition, and the BIC curves for all train passing conditions corresponding to all sensors were averaged to get the two curves shown in **Fig. 7**. The results in **Fig. 7** show that the optimal order of $m$ is 94.

In this study, CB value is finally calculated based on the Gaussian inverse cumulative distribution function (with a significance level of 1% was defined) using all baseline data sets, and then determine whether the structure is damaged by comparing CB and DF value. In addition, when the train speed is determined, the different train weight, vertical track irregularity and noise level will affect the determination of CB value. Therefore, add the white noise to the strain response of bridge to simulate a contaminated measurement response **[41]** gives:

$$\boldsymbol{b}_n = \boldsymbol{b} + nlev \times \frac{1}{N}\sum_{i=1}^{N}|\boldsymbol{b}_i| \times randn(size(\boldsymbol{b})) \tag{13}$$

where $b_n$ and $b$ are the noise-contaminated and noise-free signals, respectively. $N$ is the length of the vector $b$, *nlev* is the noise level, and *randn* is the function that generates a standard normal distribution vector.

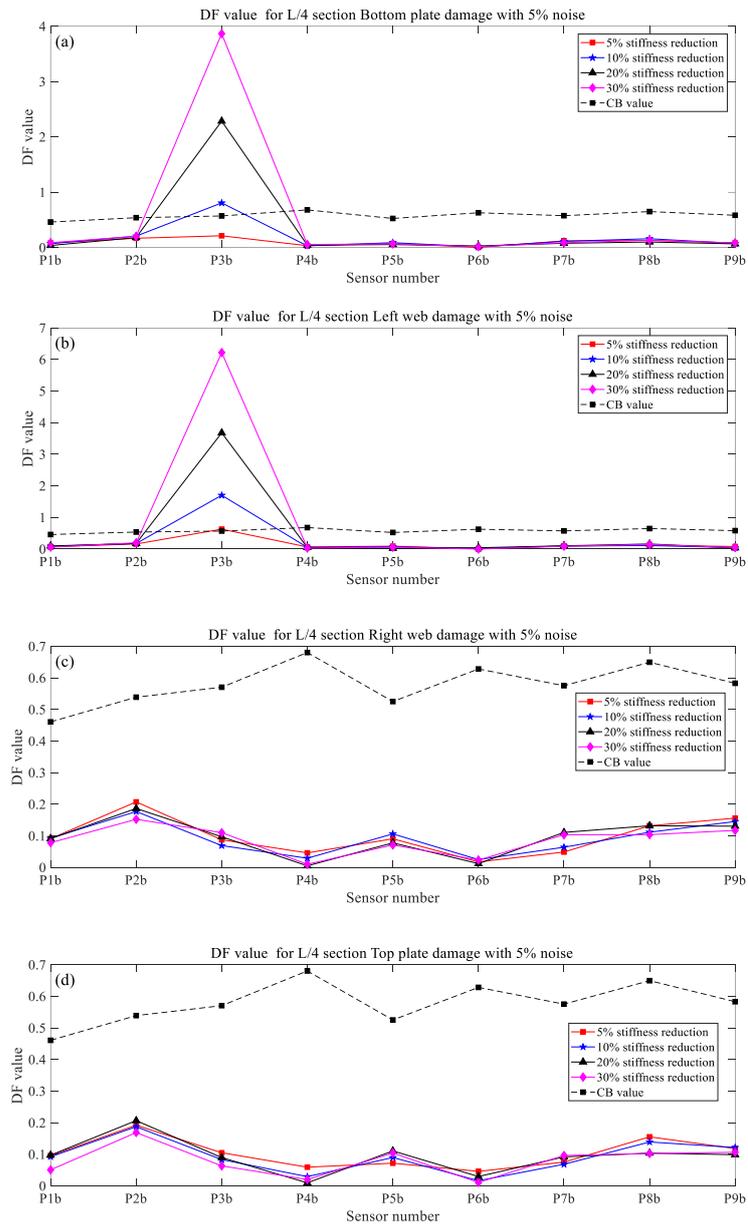

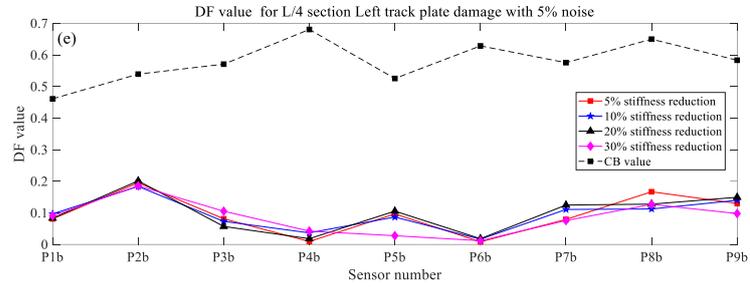

**Fig. 8.** Identification results of L/4 section when P1b to P9b are used for damage detection

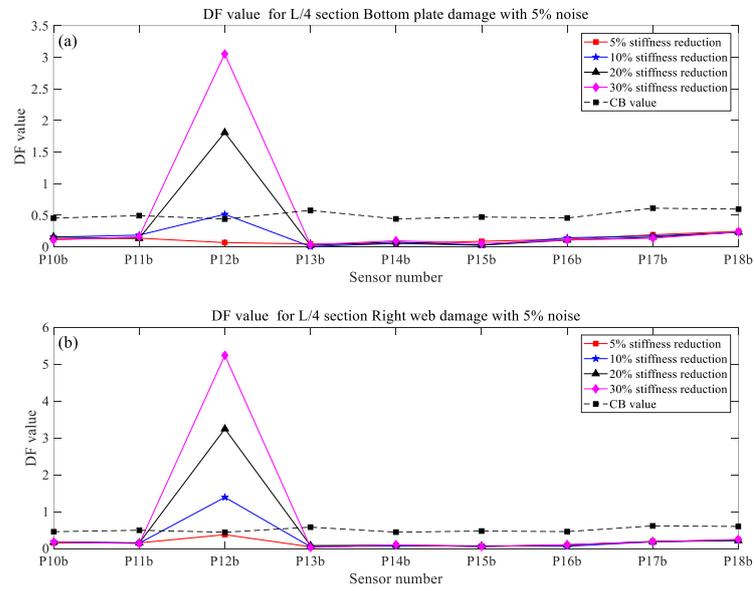

**Fig. 9.** Identification results of L/4 section when P10b to P18b are used for damage detection

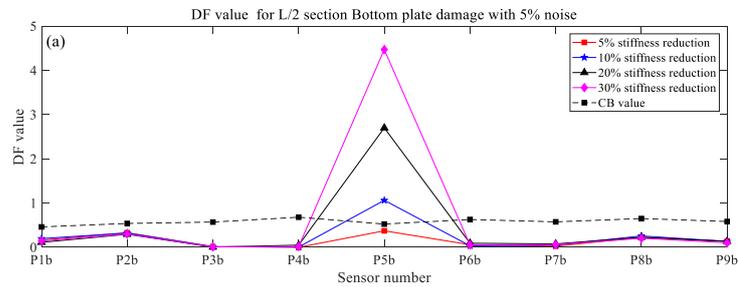

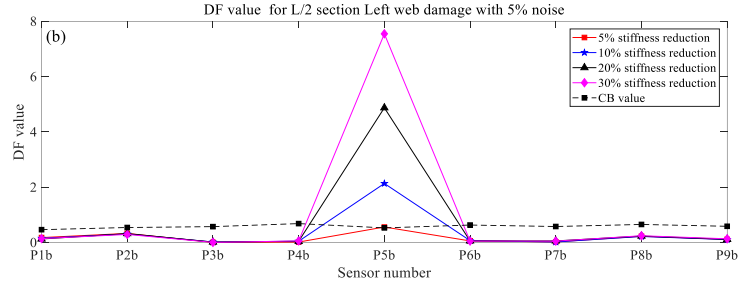

**Fig. 10.** Identification results of L/2 section when P1b to P9b are used for damage detection

When the noise level is 5% and the passing train is on the up-side lane as shown in **Fig. 1**. **Fig. 8** shows the identification results of each component of the L/4 section when sensors P1b to P9b are used for damage detection. As can be seen from **Fig. 8**, when the stiffness reduction coefficient reaches 10%, the damage of the bottom plate and the left web can be detected at point P3b (L/4 section). With the increase of the stiffness re-duction coefficient, DF value also increases. In addition, the point P3b could not detect the damage of the right web, top plate and track plate, because it was located on the bottom plate and the left web at the same time, and was far away from other components.

**Fig. 9** shows the identification results of components of the L/4 section when sen-sors P10b to P18b are used for damage detection. Unlike in **Fig. 8**, P12b can detect damage to the right web because it is located below the right web. In addition, when points P3b and P12b are used to detect the damage of the bottom plate at the same time, the point P3b can detect smaller damage. This is because the train is closer to point P3b when it passing through the bridge on the up-side lane as shown in **Fig. 1**, and point P3b has more sensitive dynamic response under the bent-torsional coupling effect.

**Fig. 10** shows the identification results of components of the L/2 section when sensors P1b to P9b are used for damage detection. Under the same train operating conditions, compared with **Fig. 8**, for the same stiffness reduction factor of the same component, **Fig. 10** has a larger DF value. The reason is that the dynamic response of L/2 section is larger than that of L/4 section for the same component, that is, point P5b has more sensitive dynamic response compared with P3b.

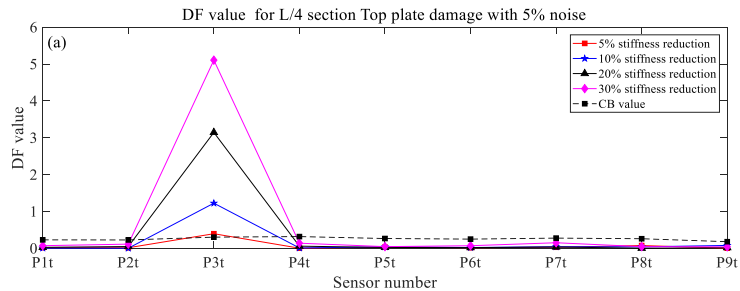

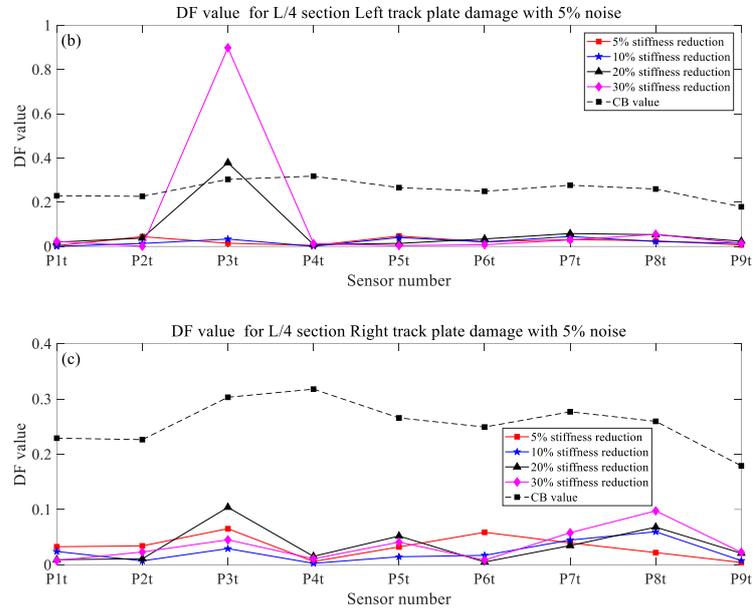

**Fig. 11.** Identification results of L/4 section when P1t to P9t are used for damage detection.

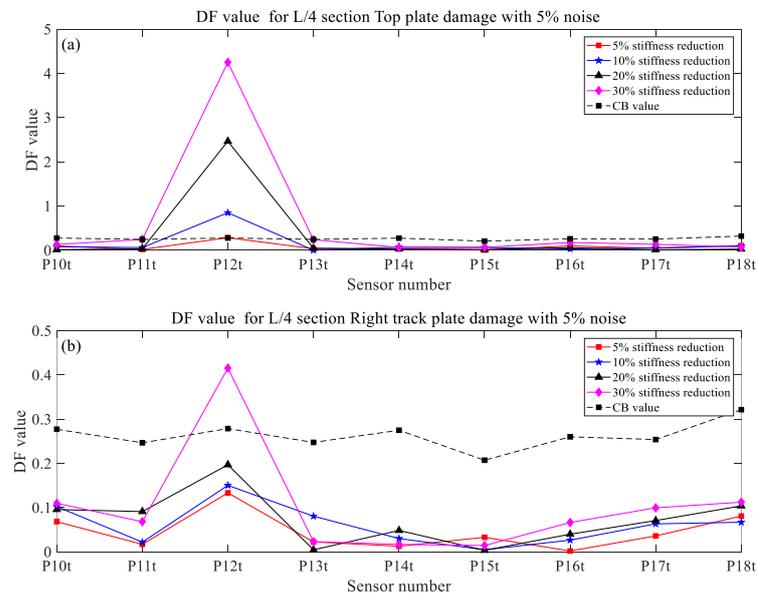

**Fig. 12.** Identification results of L/4 section when P10t to P18t are used for damage detection.

**Fig. 11** and **Fig. 12** show the identification results of some components of the L/4 section when sensors P1t to P9t and P10t to P18t are used for damage detection, respectively. As can be seen from **Fig. 11** and **Fig. 12**, when the stiffness reduction coefficient reaches 10%, the damage of the top plate can be detected at the points P3t and P12t (L/4 section). Similarly, with the increase of the stiffness reduction coefficient, DF value also increases. In addition, the point P3b could not detect the damage of the right track plate, though it was located on the top plate, and was far away from right track plate in **Fig. 1**.

In addition, although P3t and P12t are not located on the track plate, P3t and P12t are close to the left track plate and the right track plate respectively. Therefore, when the stiffness reduction coefficient is large enough, P3t and P12t can detect the damage in the left track plate and the right track plate respectively. Furthermore, the train passing through the bridge on the track near point P3t, so the point P3t has more sensitive dynamic response under the bent-torsional coupling effect.

In general, when the identified section is installed with four sensors as shown in **Fig. 1**, the damage of the six components of this section can be successfully identified, located and quantified using the proposed method in this paper, regardless of whether the train goes up or down through the bridge.

### 3.4 Effect of different vertical track irregularity, train weights and speeds

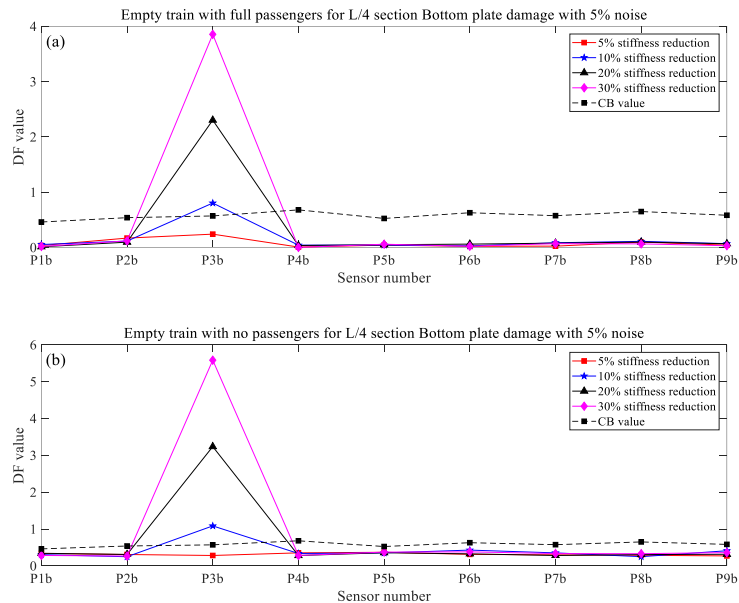

**Fig. 13.** Damaged cases for L/4 section with different train weights

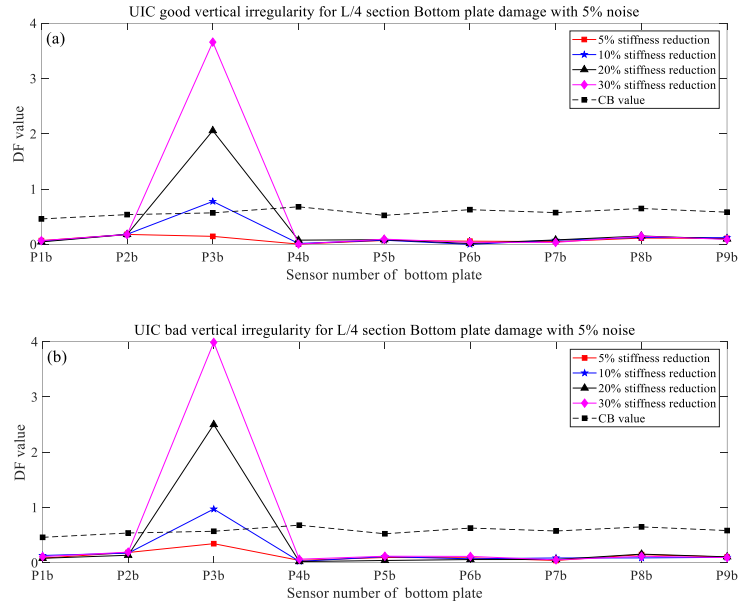

**Fig. 14.** Damaged cases for L/4 section with different vertical track irregularity

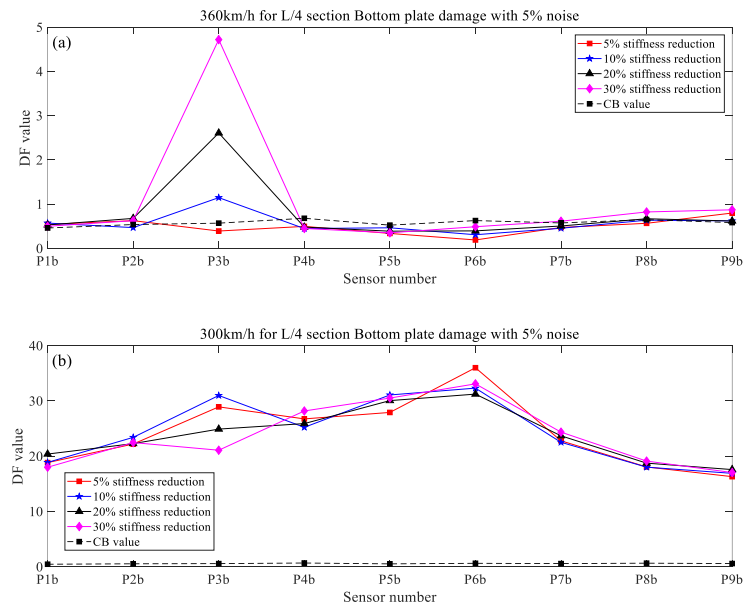

**Fig. 15.** Damaged cases for L/4 section with different train speeds

For the damage identification of bottom plate, as shown in **Fig. 13** and **Fig. 14**, when the train speed is 360km/h and the noise level is 5%, under different train weights and vertical track irregularity conditions, when the stiffness reduction factor of components reaches 10%, the proposed method in this paper can effectively identify, locate and quantify the damage.

When the undamaged condition with a speed of 360 km/h is taken as the baseline data, and the damaged condition with a speed of 360 km/h and 330 km/h are taken as the test data set respectively. **Fig. 15** shows that the damage detection, location and quantification of the condition at 360km/h have been successfully identified, but the condition at 330km/h has failed. The results show that the proposed method in this paper is sensitive to train speed.

Fortunately, in the SHM system, the train speed can be accurately identified by using sensors such as radar velocimeter. In addition, some algorithms **[43]** for identifying the accurate train speed based on dynamic strain response have been successfully applied. Therefore, in the high-speed railway health monitoring system, when the new data is obtained, the train speed can be identified first with a radar velocimeter or the speed identification algorithm in reference **[43]**, and then the damage can be successfully identified, localized and quantified using the proposed method.

## 4. Conclusions

This paper proposes a damage detection method based on the high-speed railway train-induced signals considering the bending and torsion coupling effect of box girder. The following conclusions could be drawn:

- Under the coupling effects of bending and torsion, the deformation of the box girder does not conform the Euler beam bending theory. The global response of the box girder cannot reflect the local damage characteristics, the traditional damage detection method based on the Euler beam theory cannot be applied. In this paper, the box girder section is divided into different components based on the plate element analysis method. The proposed method in this paper can successfully identify the potential damage of all components for a section by using 4 strain sensors.
- The numerical simulation results show that the proposed method in this paper is sensitive to train speed, so the train speed need to be calculated first by using a radar speedometer or the related algorithm. Even under effect of noise, different train weights and vertical track irregularity conditions, when the stiffness reduction factor of components is large enough, the proposed method in this paper can effectively identify, locate and quantify the damage.

The proposed method can effectively identify the potential damage of all components by using 4 strain sensors for the key section using train-induced responses, which is helpful for developing effective maintenance strategies for high-speed railway box girder. It is of certain practical significance for the establishment of high-speed railway health monitoring and safety early warning system.

## Acknowledgements

Financial support for this study was provided by China Railway Design Corporation R&D Program [2020YY340619, 2020YY240604], and Fundamental Research Funds for the Central Universities [FRFCU5710051018].